\documentclass[runningheads]{llncs}
\usepackage[T1]{fontenc}
\usepackage{graphicx}
\usepackage{amsmath,amsfonts,amssymb}
\usepackage{graphicx}
\usepackage{multirow}
\usepackage{lscape}
\usepackage{blindtext}
\usepackage[colorlinks=true, allcolors=black]{hyperref}
\usepackage{booktabs,caption}
\usepackage[flushleft]{threeparttable}
\usepackage{graphicx}
\usepackage{float}
\usepackage[margin=42mm]{geometry}                        
\usepackage[utf8]{inputenc}                               
\usepackage{url}                                          
\usepackage{booktabs}                                     
\usepackage{multirow}                                     
\usepackage{amsfonts}                                     
\usepackage{amsmath}                                      
\usepackage{nicefrac}                                     
\usepackage{microtype}
\usepackage{authblk}
\usepackage{booktabs}
\usepackage{graphicx}
\usepackage[numbers]{natbib}
\begin{document}

\title{Improving Explanations: Applying the Feature Understandability Scale for Cost-Sensitive Feature Selection}
\titlerunning{Application of the Feature Understandability Scale}

\author{
Nicola Rossberg\inst{1,2}\orcidID{0009-0005-3883-5833} \and
Bennett Kleinberg\inst{4}\orcidID{0000-0003-1658-9086} \and
Barry O'Sullivan\inst{1,2,3}\orcidID{0000-0002-0090-2085} \and
Luca Longo\inst{1,2,3}\orcidID{0000-0002-2718-5426} \and
Andrea Visentin\inst{1,2,3}\orcidID{0000-0003-3702-4826}
}

\authorrunning{N. Rossberg et al.}

\institute{
School of Computer Science and Information Technology,
University College Cork, Ireland
\and
Centre for Research Training in Artificial Intelligence,
University College Cork, Ireland\\
\email{n.rossberg@cs.ucc.ie}
\and
Insight Centre for Data Analytics,
University College Cork, Ireland
\and
Department of Methodology and Statistics,
Tilburg University}

\maketitle

\begin{abstract}
With the growing pervasiveness of artificial intelligence, the ability to explain the inferences made by machine learning models has become increasingly important. Numerous techniques for model explainability have been proposed, with natural-language textual explanations among the most widely used approaches. When applied to tabular data, these explanations typically draw on input features to justify a given inference. Consequently, a user’s ability to interpret the explanation depends on their understanding of the input features. To quantify this feature-level understanding, Rossberg et al. introduced the Feature Understandability Scale \cite{rossberg2025feature}. Building on that work, this proof-of-concept study collects understandability scores across two datasets, proposes a co-optimisation methodology of understandability and accuracy and presents the resulting explanations alongside the model accuracies. This work contributes to the body of knowledge on model interpretability by design. It is found that accuracy and understandability can be successfully co-optimised while maintaining high classification performances. The resulting explanations are considered more understandable at face value. Further research will aim to confirm these findings through user evaluation. 

\keywords{Psychometrics  \and Machine Learning \and Interpretability by Design \and Explainable Artificial Intelligence \and Evaluation Methods}
\end{abstract}

\section{Introduction}

In recent years, machine learning has become increasingly pervasive across a wide range of domains, including impactful areas such as medicine and agriculture, and its potential impact is undeniable \cite{phillips2012ai, pallathadka2023impact, bao2023revolutionizing, sharifani2023machine}. However, the risks associated with implementing black-box models, especially in critical fields such as medicine, must be acknowledged. The ability to audit machine learning systems is crucial to prevent the deployment of inequitable models, resulting from biased data or poor training. To promote the auditability of black- and grey-box machine learning models, a wide range of explainable AI (XAI) techniques has been proposed \citep{wang2024roadmap}. 
Through the implementation of XAI techniques, model bias can be identified and mitigated, and the legal compliance and safety of machine learning models ensured \citep{longo2024explainable}.\\

One common form of explanation is textual, in which the model's inferences are expressed in natural language \citep{vilone2021classification}. This is particularly advantageous for tabular data, where a model's output is explained in terms of its input features. For example, when applying for a loan, the input features may include 'income' and 'character'. If the loan were rejected, the textual explanation may state that ``the loan was rejected due to `income' being below 40,000 and `character' being too low". Textual explanations are generally intuitive. However, their utility depends on users' ability to understand the features they include. In this example, the explanation is accessible only if the user understands the concepts of 'income' and 'character'. As such, it is in the system designer's best interest to prioritise understandable features in explanations to maximise their efficacy.\\

To quantify feature understandability, Rossberg et al. \cite{rossberg2025feature} proposed the \textit{Feature Understandability Scale} (FUS), which measures feature understandability across various dimensions using user input. A final understandability score is assigned to each feature by averaging its ratings. To provide tailored explanations for the given user base, the scale should be completed by a subsample of end users to gauge the average understanding within the cohort. The understandability scores can then be used to co-optimise explanation quality and accuracy during machine learning model training. By favouring more understandable features during model design, the resulting explanations are expected to be more accessible and, consequently, more useful to the end user. To test this hypothesis and operationalise the scale, this proof-of-concept study implements the FUS to collect understandability ratings across two datasets, co-optimises the scores with accuracy, and assesses changes in the quality of the resulting explanations. To this end, three research questions are stipulated:

\begin{enumerate}
    \item What are the trends in the distribution of FUS scores across two datasets?  
    \item How can accuracy and understandability be co-optimised in the machine learning workflow?
    \item How does co-optimisation affect accuracy and explanation quality?
\end{enumerate}

To answer these questions, a brief literature review of existing XAI methods and cost-sensitive feature selection is conducted in Section \ref{sec: Literature}. Subsequently, user ratings of two datasets are solicited using the FUS. The datasets were selected based on their domain relevance, feature types and perceived understandability of features. Subsequently, the understandability scores are used to co-optimise accuracy and understandability. This process is detailed in Section \ref{sec: Methodology}. The final explanations are compared with those produced by a machine learning model optimised solely for accuracy, and the success of the proposed method is evaluated in Section \ref{sec: Results and Discussion}.

\section{Background} \label{sec: Literature}

\subsection{Explainable Artificial Intelligence}

The pervasiveness of machine learning models across a range of domains has increased considerably over the past few years. However, despite the increase in processing power and the high accuracy of these models, the dangers associated with their black-box nature, including biases and shortcut learning, pose a significant barrier to their widespread implementation \cite{bassi2024explanation, liu2025improving, simuni2024explainable, shah2023enhancing}. One approach to address these concerns is to implement XAI techniques, which enable the audit of decision processes and have been shown to promote model transparency and trust \citep{wiratsin2025effectiveness}. To address the breadth of existing model architectures and provide a range of explanation types, a large number of XAI techniques have been proposed, including counterfactuals \cite{kuhl2023better,baron2023explainable,hollig2023semantic}, feature attribution \cite{scholbeck2023algorithm, koenen2024toward}, latent-space mechanistic interpretability \cite{Ahmed2022,zimmermann2023scale,kastner2024explaining,ahmed2025latent} and natural language methods \cite{danilevsky2021explainability,zini2022explainability, Vojtech2026}. A full review and taxonomy of XAI methods can be found at Vilone and Longo \cite{vilone2021notions}. \\

This study focuses on the use of textual natural-language explanations for tabular data. Textual explanations are advantageous because they are easily accessible and explain decision processes by referring to real variables. However, they also increase the potential for exaggerated user trust and confidence, as well as the anthropomorphisation of the algorithm, and therefore need to be situated correctly within the given context \citep{devrio2025taxonomy, placani2024anthropomorphism, xuan2025comprehension, brdnik2025non}. Another drawback of textual explanations that has not previously been considered is that users' understanding of the explanation as a whole depends on their understanding of the input features it includes. Linking back to the previous example, a user's ability to understand why their loan was rejected depends on their understanding of the features `income' and `character'. To promote the use of understandable features, cost-sensitive feature selection approaches enable the co-optimisation of understandability and accuracy during machine learning training. Understandability scores are measured using the FUS \cite{rossberg2025feature}. The scale assesses feature-level understandability across two dimensions (`Feature-Outcome Relation' and `Understanding and Measurement') and 8 or 9 items for numerical and categorical items, respectively. By deploying the scale to a sample of the explanation's target population, an average understandability score can be computed for each feature, and its understandability cost quantified.

\subsection{Cost-Sensitive Feature Selection}
Cost-sensitive feature selection incorporates cost into the feature selection framework, where cost is either the test cost or the misclassification cost of including the feature \cite{benitez2019cost, liu2017cost, bian2016efficient}. Costly misclassifications are domain-specific and user-specified. An example is the misclassification of cancerous cells as benign, which is more costly to public health than misclassifying healthy cells as cancerous, and features may be selected to promote false positives over false negatives \cite{zhou2016cost}. Test costs refer to the cost of feature acquisition, for example, through medical tests, where a more expensive test may yield a more accurate result \cite{jiang2019wrapper}.

In the literature, three types of cost-sensitive feature selection methods are distinguished: filter methods, wrapper methods and embedded methods \cite{zhao2019cost}. 
Filter methods pre-select features before model training, and feature cost is used as a scoring function. Features are then selected based on a user-defined threshold or feature number, selecting only the N cheapest features. Filter methods are often advantageous as they are computationally cheap and can be applied to high-dimensional datasets \cite{bach2017cost}. Wrapper methods use the performance of the machine learning model to evaluate feature selection. As such, feature subsets are evaluated by the same model, which is subsequently used for classification. 
This is advantageous, as the selected features are adapted to the learner and are therefore more accurate \cite{bach2017cost}. However, they are also more computationally expensive \cite{jiang2019wrapper}. An example of this is the recursive feature elimination used in support vector machines \cite{guyon2002gene, zhao2019cost}. Both the filter and wrapper methods treat feature selection and classification as distinct steps. Embedded methods, on the other hand, combine feature selection and learner training within a single process. Feature cost is integrated into the machine learning model's loss function and optimised alongside performance during training \cite{ma2024class, gao2018class}. A summary of the three methods for cost-sensitive feature selection alongside their respective benefits and drawbacks is presented in Table \ref{tab: CSFS}.\\

\renewcommand{\arraystretch}{1.2}
\begin{table}[]
\centering
\caption{Summary of the benefits and drawbacks of the three cost-sensitive feature selection methods.}
\label{tab: CSFS}
\resizebox{\columnwidth}{!}{%
\begin{tabular}{llll}
\hline
 &  & $+$ & $-$ \\ \hline
Filter & Select features before training & \begin{tabular}[c]{@{}l@{}}Computationally cheap\\ Good for high-dimensional datasets\end{tabular} & \begin{tabular}[c]{@{}l@{}}Requires prior knowledge of costs\\ Ignores feature interaction\end{tabular} \\ \hline
Wrapper & Select features through model performance & \begin{tabular}[c]{@{}l@{}}Features adapted to learner\\ Accounts for feature interactions\end{tabular} & \begin{tabular}[c]{@{}l@{}}More computationally expensive\\ Model-specific\end{tabular} \\ \hline
Embedded & Integrate feature cost into loss function & Better performance & Not suitable for imbalanced data \\ \hline
\end{tabular}%
}
\end{table}

The current work differs from the traditional literature on cost-sensitive feature selection in that feature costs are user-defined rather than error- or test-based. Where traditionally the feature cost is defined through its contribution to a costly misclassification or the cost of acquisition (e.g. the cost of a given medical test), feature cost here is operationalised as the inverse of understandability, stipulating that a costly feature is one that detracts from the understandability of the final explanations \cite{liu2017cost, zhou2016cost, bian2016efficient}. Additionally, the success of the final model is not assessed by misclassification rates (i.e., false positives and false negatives), but rather by the quality of the final explanation as evaluated by the target audience. Cost is operationalised as the inverse of the feature's understandability measured by the FUS. As such, a more costly feature is less understandable to the end-user; hence, understandable (and cheaper) features are prioritised during the cost-sensitive feature selection procedure. This study aims to minimise this cost by co-optimising it alongside accuracy. Unlike traditional literature, success is assessed by the final understandability of explanations and the accuracy of the co-optimised model. 
\subsection{Explanation Evaluation}

The evaluation of explanation quality has been investigated in past literature, and several different approaches have been proposed \cite{zhou2021evaluating, balog2020measuring, yang2019evaluating, mohseni2021quantitative, domnich2025predicting}. Dai et al. \cite{dai2022fairness} investigated how disparities in explanation quality may lead to misplaced trust and perpetuated bias and proposed a metric to evaluate disparities in explanations given a protected attribute. However, while this permits the evaluation of disparity within the model, it does not include a human-in-the-loop approach to evaluate the impact of the final explanations. Including human evaluation of explanation quality is helpful, as the perceived utility of explanations is user-dependent and the quality of an explanation should be evaluated by the group towards which it is targeted \cite{schuff2022challenges, hilton1990conversational}.

The properties of a good explanation are user and context-dependent. In general, it is important that the explanation be on the right level and include sufficient detail for the targeted user group. The length of the explanation should also be tailored to the usage scenario, where a lay-user may have less time to read an explanation, a practitioner or data scientist may be willing to spend the time parsing a longer explanation \cite{chen2022makes}. When decomposing explanations to their purpose, Halpern et al. \cite{halpern2005causes} propose that a good explanation provides an answer to a 'why' question. In addition, Hoffman et al. \cite{hoffman2018metrics} state that a good explanation permits an a priori judgement regarding its quality and hence must be both precise and clear. While these metrics permit intermittent evaluation of explanations through qualitative analysis, the final assessment of an explanation should be based on user feedback. 

Such user-based evaluations should be collected consistently and reproducibly. In the field of psychometrics, this is possible through the development and validation of scales that capture user attitudes reliably \cite{carpenter2018ten}. Vilone and Longo \cite{vilone2023development} proposed such a scale to measure the quality of machine-generated rule-based explanations, in which the user is queried about the clarity they have gained from the explanations. In a similar vein, Holzinger et al. \cite{holzinger2020measuring} proposed the system causability scale, which measures the quality of a human-AI interface or an explanation process. This allows the user to assess the extent to which the communication method makes the provided explanations accessible. While a range of explanation evaluation methods is available, none have been developed for analysing natural language explanations of tabular data. As this study aims to test the collection and co-optimisation of understandability scores, the quality of explanations will be evaluated through changes in understandability scores and qualitatively through example explanations. 

\subsection{Explainability-Accuracy Trade-off}

The existence of an explainability-accuracy trade-off is contentious in the literature. On the one hand, it is argued that the implementation of more explainable grey-box models decreases accuracy due to their associated simplicity, and vice versa, complex black-box models lead to high performance but low interpretability \cite{assis2023explainable, hunsicker2025investigating, shuvra2024explainability, kabir2025review, assis2025performance}. The opposing argument states that this trade-off, while theoretically sound, is seldom observed in practice, in large part due to the lack of objective measurement of interpretability, and more transparent models should hence be favoured by default \cite{bell2022s, dziugaite2020enforcing, carrington2018measures}. Additionally, user preferences and understanding of models may be domain-dependent, introducing a further consideration for model design \cite{hunsicker2025investigating}. In addition to the inherent interpretability of a model, a range of post-hoc explainability methods such as GradCAM, SHapley Additive exPlanations (SHAP) and LIME have been proposed, which can provide explanations for model behaviour without impacting classification accuracy \cite{selvaraju2017grad, lundberg2020local2global, ribeiro2016should}. However, some of these methodologies may be less faithful to the system process, as they provide explanations through proxy models. As such, several important considerations need to be taken into account during model design to ensure suitable classification and explanation performance for the given task.

\subsection{The Feature Understandability Scale}

The FUS was proposed and validated by Rossberg et al. in \cite{rossberg2025feature}. The scale is a validated measurement instrument that quantifies the user's self-assessed understanding at the feature level. The FUS is divided into two subscales: one for categorical features and one for numerical features, and users complete the respective scale for each feature in a dataset. Users are defined as the target audience of the machine learning explanations based on the dataset. The scale is evaluated on a five-point Likert scale, and the mean rating is computed as the understandability score for each feature.

\section{Methodology} \label{sec: Methodology}

The following details the data collection and co-optimisation of the understandability scores. This research aimed to assess the distribution of understandability scores across two domains and to implement a new methodology for their co-optimisation with accuracy. The overarching goal is to generate improved explanations on an ante-hoc basis. The section is structured according to the first three steps of the CRISP-DM guidelines \cite{wirth2000crisp}. These guidelines standardise the structure of data mining projects, promoting clear communication of the research methodology employed.

\subsection{Business Understanding}
\begin{figure}
    \centering
    \includegraphics[width=\linewidth]{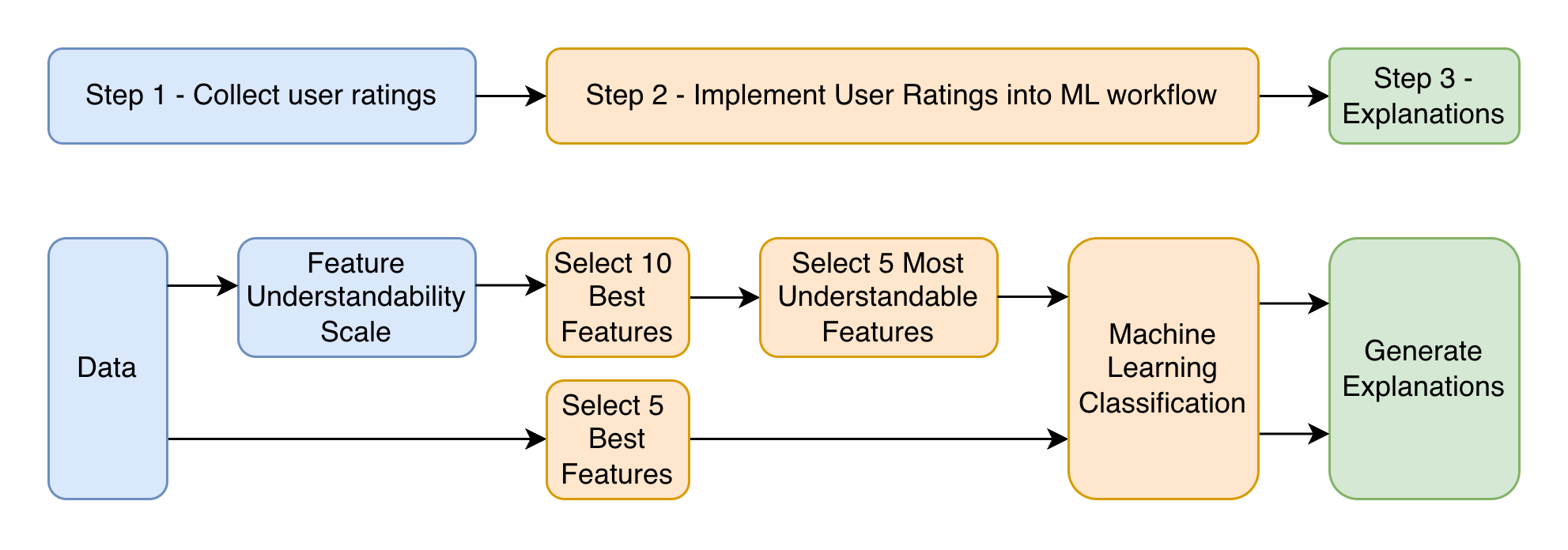}
    \caption{Workflow of this study.}
    \label{fig: workflow}
\end{figure}

This study aims to assess how understandability scores are distributed, how they can be integrated into machine learning training, and whether they enable the generation of more understandable explanations. From an end-user perspective, explanations are considered useful if they are both understandable and at the appropriate level \cite{verdejo2011levels}. 
As such, feature understanding should be adjusted to the specific target group of the explanations. For example, medical explanations targeted at patients need to differ from those targeted at practitioners. As such, the FUS should be completed by the target group. 
As a proof-of-concept, this work aims to test whether the implementation and co-optimisation of scores are feasible and beneficial, and therefore did not restrict participants to the target cohort. 
The success of the proposed methodology is evaluated by inspecting the feature understandability scores of both the traditional and co-optimised explanations, as well as the accuracy of the respective models. 
This aims to show both the degree to which understandability improves and the overlap between important and understandable features. 
The project will be structured in three key phases. First, two datasets from different domains are selected, and the understandability scores of their features are acquired. 
Second, accuracy and understandability are co-optimised, and both traditional and improved explanations are generated.
Finally, the generated co-optimised explanations are qualitatively evaluated and compared to the traditional explanations. The exact workflow of the paper is showcased in Figure \ref{fig: workflow}. The figure shows the collection of understandability ratings in the first step of the study. During the co-optimisation procedure, the ten best features are selected using traditional feature selection; of those, the five most understandable are retained and classified using machine learning models. In the traditional feature selection comparison, the five best features are selected directly from the dataset and used to classify. Local explanations for both models are generated, and the accuracy and understandability scores are presented. \\

\subsection{Data Understanding}
The FUS is applied to two public datasets to ensure that any changes in explanation quality are not domain-specific. The first dataset is the `Telco Customer Churn' dataset, which contains data that can be used to predict whether a customer will switch service providers given their family and usage history. 
The dataset is publicly available \footnote{\url{https://www.kaggle.com/datasets/blastchar/telco-customer-churn}}. This dataset will be referred to as `Dataset 1' going forward. The second dataset is the `Body Signals of Smoking' dataset and contains data useful for the prediction of whether a patient is a smoker, given a range of physiological indicators. The dataset is publicly available \footnote{\url{https://www.kaggle.com/datasets/kukuroo3/body-signal-of-smoking}} and will be referred to as `Dataset 2' going forward. Before collecting the understandability scores, the datasets were cleaned by removing features with zero variance. Furthermore, user ID variables were removed because they were unique patient identifiers and lacked predictive relevance.

Understandability scores were collected using a survey on the platform Qualtrics\footnote{\url{https://www.qualtrics.com/}}. Data collection was approved by the University College Cork Social Research Ethics Committee under Log Number 2025-231 on the 1st of October 2025.
Participants were recruited via convenience and snowball sampling, and no personal data were collected. Snowball sampling was used by asking participants to pass the survey to their networks, thereby exponentially increasing the reach of the data collection. Participants had to be over 18 and have a self-assessed English level of B1 or higher. All participants had to provide informed consent before commencing the study.
Participants were assigned to one of the two datasets and asked to rank half of the features using the FUS. Each participant ranked 10 (Dataset 1) or 12 (Dataset 2) features using the scale. Participants were asked to rate only half the features to limit workload and avoid respondent fatigue. Features were presented in a pseudo-random order and counterbalanced to ensure that each feature received an equal number of ratings. Participant responses were cleaned, and fast responders were removed from the dataset. Additionally, all responders with a self-reported English Level below B1 were removed. The understandability scores were then reverse-coded to yield cost values ranging from 1 (lowest cost, most understandable) to 5 (highest cost, least understandable). Where previously more understandable features were given higher ratings, the reverse coding reverses the scale, assigning a lower cost to more understandable features and vice versa. This is done to align the methodology with the existing literature on cost-sensitive feature selection. As such, more costly (and less understandable) features are considered more expensive to the acquisition process and cheaper (more understandable) features are favoured. Finally, the mean understandability cost was computed for each feature. 

\subsection{Modelling}
After data collection, the understandability cost scores were integrated into the machine learning pipeline using a filter method, enabling the creation of improved explanations on an ante-hoc basis. A filter method was selected for the co-optimisation procedure to ensure transparent feature selection, permit replication and minimise confounding effects that detract from the goal of the current research. First, a preliminary model selection was conducted using stochastic resampling to test the implementation and evaluate the performance of a Decision Tree, a Random Forest, and a Support Vector Machine (SVM) on the training set, with the models retained based on average classification performance. For each model, Select K Best with chi-square, f-classif, and mutual-info-classif, as well as Recursive Feature Elimination, were tested and retained based on the highest cross-validation accuracy in the training set. Chi-Square computes the exponential chi-square kernel, F-classif the ANOVA F-value, and mutual-info-classif computes the mutual information based on k-nearest neighbours estimation between each feature and the target variable. Based on these respective information scores, Select K Best ranks all features and retains the K best features, where K is user-specified. Recursive Feature Elimination is a more complex procedure in which a model (here, a Decision Tree) is trained, and the assigned feature importances are used to iteratively remove features and recompute them. All feature selection methods were implemented through Scikit-learn \cite{scikit-learn}. After selecting the machine learning and feature selection methods, accuracy and understandability are co-optimised to minimise the potential trade-off between explainability and accuracy. In the literature, it is argued that this trade-off may result from simpler models being more transparent but yielding lower accuracy levels due to their reduced complexity \citep{kabir2025review}. The current method is model-agnostic for processing tabular data and is therefore not expected to directly impact performance. However, the trade-off may result from non-understandable features being key to discriminating between classes, while understandable features may contain less information. As such, the introduction of understandability-based feature selection may lead to the exclusion of important features and the reduction of model performance. To counteract this, the proposed method aims to co-optimise accuracy and understandability, minimising predictive power loss while promoting the use of understandable features. 

\begin{figure}[h!]
    \centering
    \includegraphics[scale=0.2]{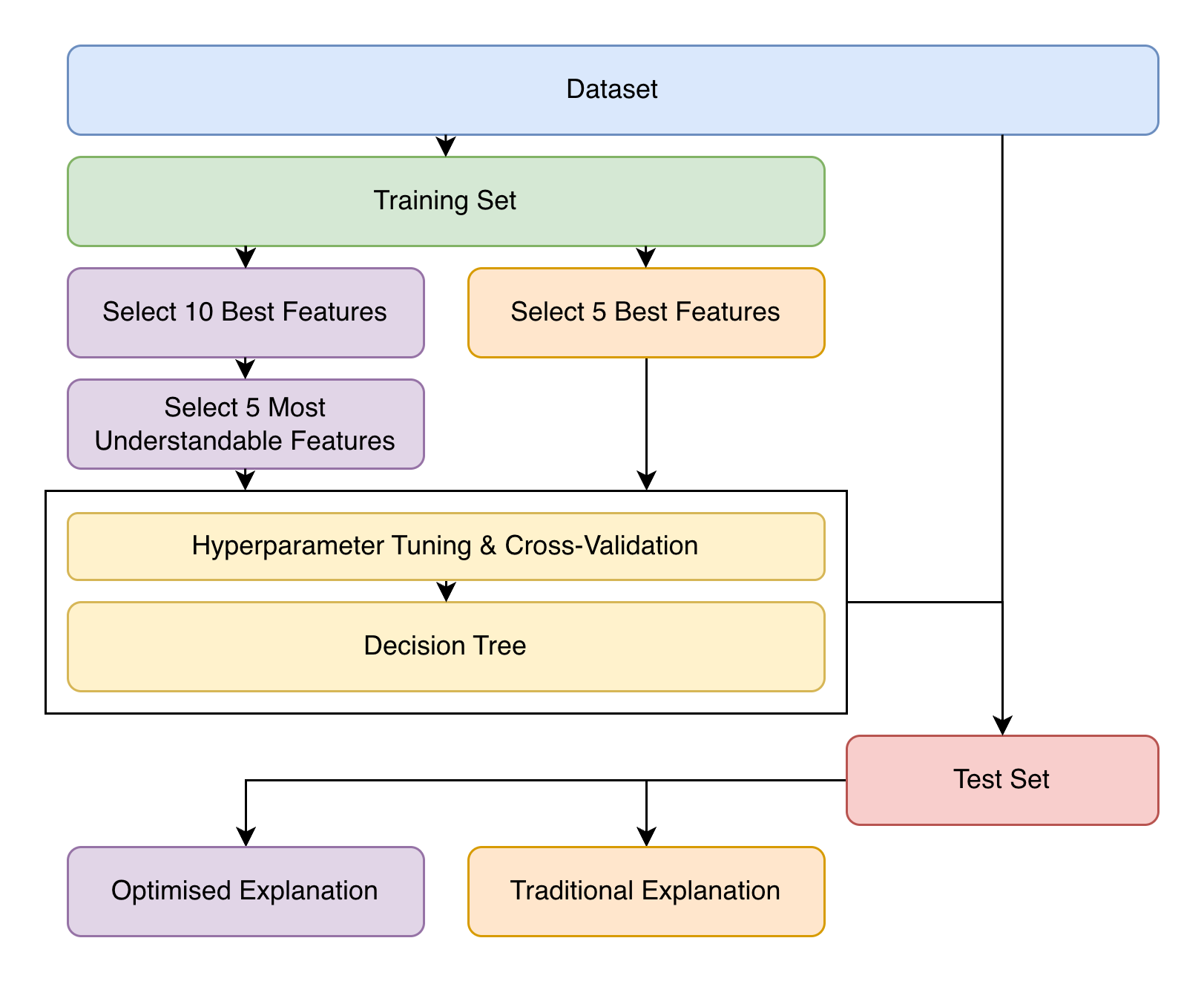}
    \caption{Workflow of the Cost-Sensitive feature selection}
    \label{fig: Co-Optimisation Workflow}
\end{figure}

The workflow of the implemented co-optimisation is depicted in Figure \ref{fig: Co-Optimisation Workflow}. The datasets were split into training and test sets with a stratified 70-30 split. The training set was further split into training and validation with a stratified 70-30 split for the model and feature selection algorithm selection procedure. Numerical features were standardised, and one-hot and label encodings were implemented for multiclass and binary categorical features, respectively. In the first step of the co-optimisation, the ten most important features were selected through traditional feature selection. Four feature selection algorithms (Select K Best with the chi-square, f-classif, and mutual-info-classif kernels, and Recursive Feature Elimination) were tested on the training data for each dataset and model, and the best-performing algorithm was retained. Of the ten most important features selected, the five most understandable were retained for classification. Three classification models (SVM, Random Forest, and Decision Tree) were tested on the training set, and the two best-performing models were implemented on the validation data. The training and validation data were then merged, and the 5-feature dataset was classified, using a grid search with 5-fold cross-validation for model tuning. Both feature selection and classification were implemented through the Scikit-learn library \citep{scikit-learn}. The best model was retrained on the entire training data and evaluated on the reserved testing set. As a comparator to the co-optimisation procedure, traditional feature selection was implemented. Here, the five most important features were selected from the training dataset using the same feature selection algorithm. A machine learning model was then trained using the same grid search and 5-fold cross-validation procedure and evaluated on the reserved test data. Due to the range of feature selection and classification algorithms tested, evaluation was conducted only using a stratified train-test split.

After model tuning and evaluation, SHAP was implemented to extract feature importances and local explanations for the chosen models \cite{lundberg2020local2global}. While some models (e.g., Decision Tree) are inherently transparent and provide more faithful ways to extract feature importances, SHAP was chosen to enable cross-model comparison. The selected SHAP model was 'Tree Explainer'. SHAP computes feature importance by averaging the change in model output when each feature is introduced one at a time, conditioning on the feature in question. The Tree Explainer is specifically tailored to tree- and ensemble-tree models, making it suitable for the current study. Feature importances were computed for a selected range of test set instances and transformed into explanations. A heuristic was created to generate examples of the final explanations. Several considerations were made during the design process, including the length of explanations, the provision of comparators for individual values, and the direction in which the individual features influenced the decision. The final explanation was structured to first present the predicted outcome, then introduce the features in favour of it, followed by those that detract from the decision. For each feature, the instance's value was provided alongside the mean (for numerical features) to contextualise the instance's position. The features in the explanations are ordered by their relative importance to the final prediction. The full code, including feature selection, model training and explanation generation, can be found at \url{github.com/ncrossberg/User-Study}.

\section{Results and Discussion} \label{sec: Results and Discussion}

\subsection{Feature Rating Collection}

\begin{figure}[h!]
    \centering
    \includegraphics[width=0.9\linewidth]{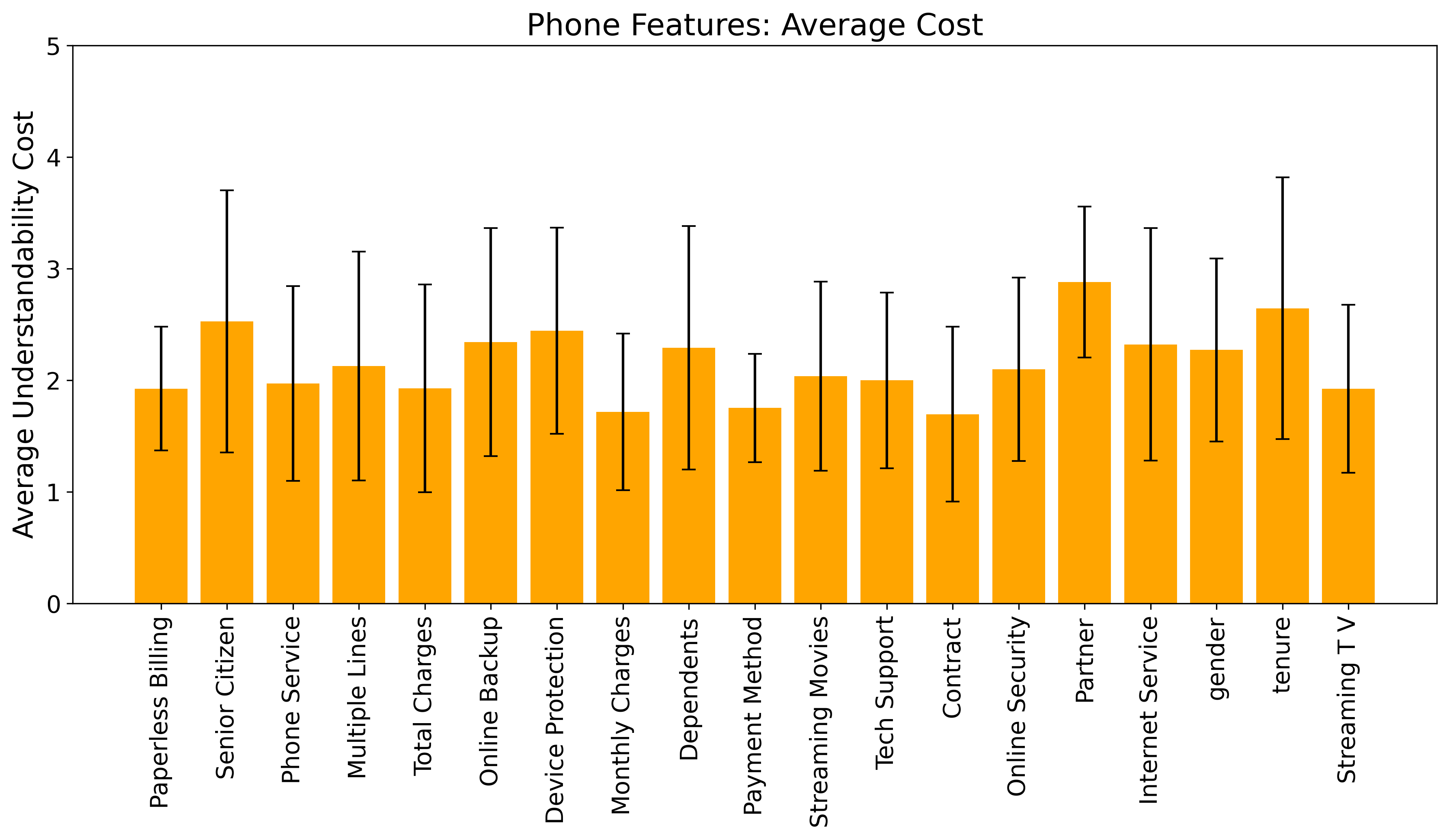}
    \caption{Average feature cost and standard deviation for the phone company customer churn dataset. Note that cost is computed as the inverse of understandability to permit co-optimisation. A lower score indicates a more understandable feature.}
    \label{fig:Phone Costs}
\end{figure}

\begin{figure}[h!]
    \centering
    \includegraphics[width=0.9\linewidth]{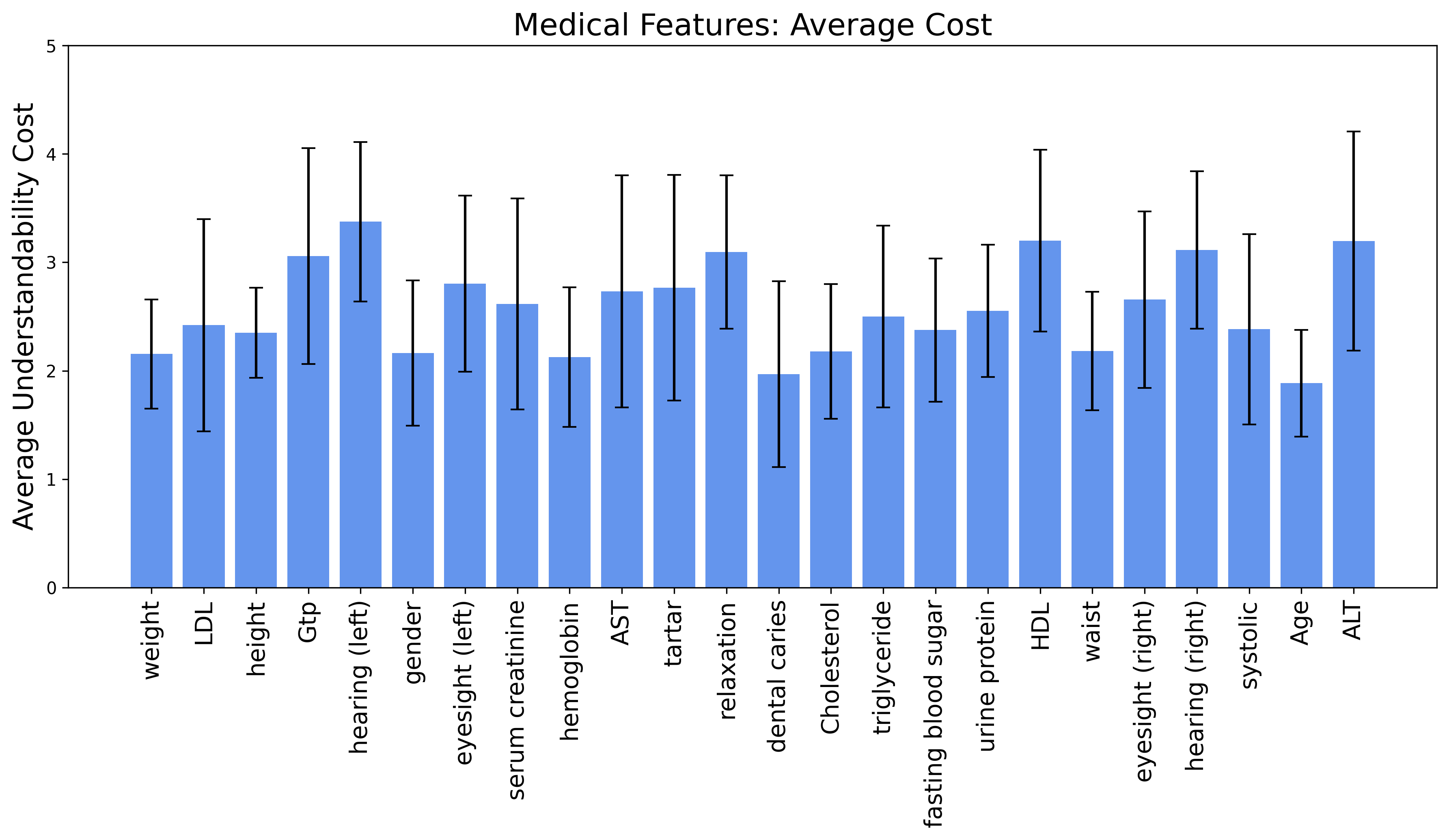}
    \caption{Average feature cost and standard deviation for the medical dataset. Note that cost is computed as the inverse of understandability to permit co-optimisation. A lower score indicates a more understandable feature.}
    \label{fig:Medical Costs}
\end{figure}

Feature ratings were collected from the 15th of October to the 1st of November 2025. A total of 163 responses were initially collected. After removing very fast respondents (less than 3 minutes) and incomplete surveys, 54 respondents remained. As each respondent rated one quarter of the features, the average number of ratings per feature was 13.53. The distributions of ratings across the two datasets, along with their standard deviations, are shown in Figures \ref{fig:Phone Costs} and \ref{fig:Medical Costs}. The distributions of ratings were rather small: Dataset 1 ranged from a minimum cost of 1.70 (Contract) to 2.88 (Partner), while Dataset 2 ranged from a minimum cost of 1.88 (Age) to a maximum of 3.38 (hearing (left)). On the other hand, the scores' standard deviation is relatively large, indicating considerable variation within the response distribution for each feature. 

Several reasons for the limited range of the results can be postulated. However, these are only theoretical assertions, and an item response or diffusion model is necessary to confirm them. One possibility is that this is an instance of 'fence sitting', a phenomenon in which respondents choose neutral answers despite having a stronger opinion. Potential reasons for fence-sitting include social desirability (claiming to understand a feature even if they don't), apathy (disinterest in the questionnaire), indecisiveness, or lack of information \cite{kulas2009middle}. While fence-sitting cannot be avoided entirely, future research may wish to include further clarifying information, as well as attention checks and financial incentives to combat this behaviour. An alternative explanation for this phenomenon is that the range of feature difficulty was, in fact, limited, and the ratings are representative of a dataset that, on average, is understandable. 
The high variance within each feature's ratings may be attributed to three potential causes. First, as a pilot study, participants were not limited to the population of interest, leading to a more diverse sample. This may cause considerable discrepancies in understanding, increasing the variance. In addition, the small sample size of this study may also increase variance. As the number of participants increases, the impact of random fluctuations decreases, and the variance of results tends to decrease. However, this is not guaranteed, as the variance may reflect a genuine discrepancy in understanding within the population of interest. A third possibility is that participants may have misunderstood the scale labels, hence introducing error to the measurements. As such, future research should clearly define the scale's endpoints and explore whether the high variance persists in larger samples. 

On average, Dataset 2 was found to be more difficult, which is in line with expectations, as respondents are likely to be less familiar with specialised medical terms. However, it is important to note that direct comparisons between the datasets are not recommended. Participants are likely to use other features within the same dataset as internal reference points when providing ratings. As a result, their interpretation of the rating scale and their assessment of feature difficulty may differ across datasets. Because each participant was exposed to only one dataset, these relative judgments are not anchored to a common baseline, preventing a meaningful direct comparison of understandability scores across datasets. Additionally, presenting a random subset of features in each dataset may contribute to the high variance. If, by chance, a participant receives only difficult features, the more understandable features would be scored as comparatively costly due to the misaligned reference. Surprisingly, the most costly feature in Dataset 2 was `hearing (left)', which was considered to be reasonably understandable at face value. However, the high cost likely stems from the measurement scale, which respondents may be unfamiliar with. A similar phenomenon is observed in Dataset 1, where 'Partner' is the most costly feature, despite being understandable at face value. One conclusion that may be drawn from this is that open, easily accessible documentation of rating scales is vital for understanding features. Future research should provide a key for the feature values during rating collection if the measurement scale is ambiguous. In addition, larger samples should be collected to improve generalisation and decrease variance by minimising the impact of outliers. The exact number of samples necessary will depend on the domain and aim of the study \cite{sinha2025sample}. 

\subsection{Co-Optimisation}
The results of the co-optimisation alongside the traditional feature selection are presented in Table \ref{tab: Quantitative Results Co-Optimisation}. After testing several feature selection and classification methods, Random Forest and Decision Tree were retained as classifiers based on their validation accuracy. The features in Dataset 1 were selected using the Select K best algorithm with an f-classif kernel for the Decision Tree, and Recursive Feature Elimination with a Decision Tree as the model for the Random Forest. Those in Dataset 2 were selected using Recursive Feature Elimination with a Decision Tree as the model for both the Decision Tree and the Random Forest. \\

The classifiers in both datasets achieve good balanced accuracies (73.08 - 75.26\% for Dataset 1 and 75.49 - 76.05\% for Dataset 2) with minimal overfitting, implying that the models are correctly specified for the data and classification task. In three out of four cases, test accuracy is slightly lower in the co-optimised compared to the traditional model. Only the Random Forest in Dataset 1 performs slightly better under co-optimisation conditions, which may be attributed to sample variations or the co-variation of understandability and feature importance. The slight drop in accuracy observed across the remaining three cases is in line with the expected accuracy-explainability trade-off, where important features may score low on understandability. Importantly, the total and mean understandability costs decrease in the co-optimised model, highlighting that the proposed method successfully selects more understandable features while retaining relatively high accuracy levels. The decrease in average cost is relatively small (0.31 (Decision Tree) and 0.3 (Random Forest) in Dataset 1 and 0.36 (Both Models) in Dataset 2. However, relative to the overall range in cost, it constitutes 26.2\% (Decision Tree) and 25.4\% (Random Forest) of the total variation in cost for Dataset 1 and 24\% (both models) for Dataset 2. As such, understandability increases considerably compared to the relative drop in accuracy, emphasising that it is possible to develop models that remain accurate while prioritising understandable features.

When comparing the two models, the Random Forest performs better in Dataset 1, demonstrating both higher accuracy values and lower feature cost. However, the model also requires significantly longer training and prediction times, and the accuracy-time-understandability trade-off should be considered on a case-by-case basis. In Dataset 2, the Decision Tree performs better in terms of accuracy, with both models selecting the same features and reporting equal understandability scores, while the Random Forest has considerably larger training and prediction times. As such, the FUS permits the consideration of not only run-time and accuracy during model selection but also the integration of understandability into this procedure.

\renewcommand{\arraystretch}{0.95}
\begin{table}[]
\centering
\caption{Balanced Accuracy (in \%) and Training Times (in seconds) for the Decision Tree and Random Forest models for Dataset 1 (customer churn) and Dataset 2 (smoking prediction)}
\label{tab: Quantitative Results Co-Optimisation}
\resizebox{\columnwidth}{!}{%
\begin{tabular}{lllll}
\hline
Decision Tree & \multicolumn{2}{c}{Dataset 1} & \multicolumn{2}{c}{Dataset 2} \\ \hline
 & Traditional & Co-Optimisation & Traditional & Co-Optimisation \\ \hline
Train Accuracy & 76.94 & 74.74 & 76.64 & 76.11 \\
CrossVal Accuracy & 76.81 & 74.74 & 76.27 & 75.99 \\
Test Accuracy & 74.56 & 73.08 & 76.05 & 75.57 \\
Total Cost & 10.76 & 9.18 & 12.8 & 11.02 \\
Mean Cost & 2.15 & 1.84 & 2.56 & 2.20 \\
Training Time & 0.0052 & 0.025 & 0.021 & 0.020 \\
Prediction Time & 0.0017 & 0.014 & 0.0033 & 0.0032 \\ \hline
Random Forest & \multicolumn{2}{c}{Dataset 1} & \multicolumn{2}{c}{Dataset 2} \\ \hline
 & Traditional & Co-Optimisation & Traditional & Co-Optimisation \\ \hline
Train Accuracy & 77.17 & 76.53 & 76.08 & 75.90 \\
CrossVal Accuracy & 77.06 & 75.92 & 75.85 & 75.85 \\
Test Accuracy & 75.05 & 75.26 & 75.7 & 75.49 \\
Total Cost & 10.31 & 8.79 & 12.80 & 11.02 \\
Mean Cost & 2.06 & 1.76 & 2.56 & 2.20 \\
Training Time & 0.81 & 0.79 & 3.14 & 2.67 \\
Prediction Time & 0.034 & 0.035 & 0.20 & 0.20 \\ \hline
\end{tabular}%
}
\end{table}

The features selected by each model across the two datasets are presented in Table \ref{tab: Feature Importances}. Two features (highlighted in bold) are selected by both the traditional and the co-optimised Decision Tree in both datasets, implying that they are both important and understandable. While it is unlikely that this pattern will generalise across all domains and datasets, the observed correlation between feature importance and understandability is remarkable. This pattern presents even more strongly in the Random Forest, where three of the same features are selected in Dataset 1. Interestingly, this selection of similar features also co-occurs with improved understandability, further supporting the correlation between informativeness and understandability in the chosen datasets. Within the domains, the features that are both important and understandable make sense intuitively. In Dataset 1, the Decision Trees select a \textit{month-to-month contract} with \textit{no tech support}. At face value, these features are both understandable and expected to be strongly correlated with customer churn, given the short-term, low-support nature of such contracts. The Random Forest also selects the \textit{month-to-month contract} alongside the \textit{monthly} and \textit{total charges} in both the traditional and co-optimised settings. As the only two numerical features in the dataset, \textit{monthly charges} and \textit{total charges} are understandable at face value and likely informative, given their strong correlation with customer loyalty (the longer a customer has stayed with a firm, the higher the total charges). Similarly, in Dataset 2, the features selected by the Decision Tree and Random Forest are \textit{gender} and \textit{age}, both of which are understandable and expected to strongly correlate with whether a patient smokes. Interestingly, the two features selected by both models in Dataset 1 show considerable differences in importance between the traditional and co-optimised models. In Dataset 2, on the other hand, \textit{gender} and \textit{age} are the two most important features in both models, further emphasising their strong predictive and explanatory relevance. 

\renewcommand{\arraystretch}{1.2}
\begin{table}[]
\centering
\caption{Feature importances for both the traditional and co-optimised feature selection methodologies for Dataset 1 (customer churn) and Dataset 2 (smoking prediction). Features are ordered from most to least important.}
\label{tab: Feature Importances}
\resizebox{\columnwidth}{!}{%
\begin{tabular}{llll}
\hline
\multicolumn{4}{c}{Decision Tree} \\ \hline
\multicolumn{2}{c}{Dataset 1} & \multicolumn{2}{c}{Dataset 2} \\ \hline
Traditional & Co-Optimisation & Traditional & Co-Optimisation \\ \hline
\begin{tabular}[c]{@{}l@{}}Tenure, \\ Internetservice: Fiber Optic, \\ Onlinesecurity: No, \\ \textbf{Techsupport: No}, \\ \textbf{Contract: Month-to-month}\end{tabular} & \begin{tabular}[c]{@{}l@{}}\textbf{Contract: Month-to-month}, \\ Contract: Two year, \\ Payment Method: Electronic check, \\ \textbf{Techsupport: No}, \\ Streaming Movies: No internet service\end{tabular} & \begin{tabular}[c]{@{}l@{}}\textbf{Gender},\\ \textbf{Age}, \\ Triglyceride, \\ ALT, \\ GTP\end{tabular} & \begin{tabular}[c]{@{}l@{}}\textbf{Age}, \\ \textbf{Gender}, \\ Waist, \\ Fasting Blood Sugar, \\ LDL\end{tabular} \\ \hline
\multicolumn{4}{c}{Random Forest} \\ \hline
Traditional & Co-Optimisation & Traditional & Co-Optimisation \\ \hline
\begin{tabular}[c]{@{}l@{}}Tenure, \\ \textbf{Monthly Charges}, \\ \textbf{Total Charges},\\ Internet Service Fiber Optic,\\ \textbf{Contract Month-to-month}\end{tabular} & \begin{tabular}[c]{@{}l@{}}\textbf{Contract Month-to-month},\\ Contract One year,\\ \textbf{Monthly Charges},\\ Payment Method: Electronic check,\\ \textbf{Total Charges}\end{tabular} & \begin{tabular}[c]{@{}l@{}}\textbf{Gender},\\ \textbf{Age}, \\ Triglyceride, \\ ALT, \\ GTP\end{tabular} & \begin{tabular}[c]{@{}l@{}}\textbf{Age}, \\ \textbf{Gender}, \\ Waist, \\ Fasting Blood Sugar, \\ LDL\end{tabular} \\ \hline
\end{tabular}%
}
\end{table}

The final generated explanations are shown in Table \ref{tab: explanations}. At face value, the features selected in the co-optimised models appear more understandable than those in the traditional model, with the included terminology being more commonly accessible. One notable aspect of the explanations in Dataset 1 for both models is the use of one-hot encoding for categorical features. Several variables in this dataset (e.g., Contract and Online Security) had multiple categories and were therefore one-hot encoded before analysis, with each encoded category retaining the same cost as the original feature. While this representation is suitable for model training, it complicates the interpretation of the resulting explanations. Specifically, the explanation may emphasise the absence of a particular category rather than the participant’s actual category membership. For example, instead of stating that a participant has a month-to-month contract, the explanation may highlight that the participant does not have a one-year contract. This framing may be unintuitive for users and may require an implicit understanding of how one-hot encoding represents categorical data. As a result, overall interpretability depends not only on the user's understanding of the original feature but also on understanding its individual categories and the relationships between them. Furthermore, presenting multiple binary indicators for a single categorical variable can fragment the information and make explanations less accessible, particularly for lay users who may not be familiar with one-hot encoding. Future research may wish to mitigate these issues by weighting the scale items that assess understanding of individual categories more heavily during FUS evaluation. Alternatively, aggregating one-hot-encoded categories before explanation creation could be explored. While this would likely make explanations more accessible, it would also decrease faithfulness, and the more suitable choice will likely be domain and context-dependent. 

In Dataset 2, the co-optimised features are somewhat more understandable, as only 2 are specialised medical terms compared to three in the traditional explanation. However, it is unclear whether the quantity or presence of non-understandable features has a greater impact on user understanding, and future research may wish to clarify this. An interesting aspect of the explanations in Dataset 2 is the use of acronyms (e.g. ALT, GTP, LDL). While acronyms may at times be more commonly known than the words they represent, they may also reduce the understandability of explanations, depending on the domain and user base. As such, it may be of interest to researchers to present abbreviations alongside the full names of variables during the understandability score collection to prevent any related ambiguity.

However, all of these hypotheses need to be verified through specific user feedback to assess the real impact of the explanation structure and presentation. In conjunction with quantitative evaluation via a scale, qualitative interviews may be useful for identifying other areas for improvement in the structure and composition of the explanations. 

\renewcommand{\arraystretch}{1.3}
\begin{table}[]
\centering
\caption{Example explanations of the co-optimised and traditional feature selection for both datasets. These explanations were created in the last step of the co-optimisation procedure through the feature importances extracted from SHAP and the designed explanation heuristic.}
\label{tab: explanations}
\resizebox{\columnwidth}{!}{%
\begin{tabular}{ll}
\hline
\multicolumn{2}{c}{Decision Tree} \\ \hline
 & Co-Optimised \\ \hline
Dataset 1 & \begin{tabular}[c]{@{}l@{}}The customer is predicted to stay with the company because their contract category is Two year and their \\ streaming movies category is No internet service. The fact that their contract category is not Month-to-month \\ and their payment method category is not Electronic check and their techsupport category is not No detracted \\ from this prediction.\end{tabular} \\ \hline
Dataset 2 & \begin{tabular}[c]{@{}l@{}}The individual is predicted to be a smoker because their gender is M and their age of 40.00 is less than average \\ and their waist of 98.00 is more than average. The fact that their fasting blood sugar of 106.00 is more than average \\ and their ldl of 106.00 is less than average detracted from this prediction.\end{tabular} \\ \hline
 & Traditional \\ \hline
Dataset 1 & \begin{tabular}[c]{@{}l@{}}The customer is predicted to stay with the company because their tenure of 6487.79 is more than average \\ and their contract category is Two-year. The fact that their online security category is not No and their \\ techsupport category is not No and their contract category is not Month-to-month detracted from this prediction.\end{tabular} \\ \hline
Dataset 2 & \begin{tabular}[c]{@{}l@{}}The individual is predicted to be a smoker because their gender is M, their age of 40.00 is less than average, \\ their gtp of 127.00 is more than average, their alt of 24.00 is less than average. The fact that their triglyceride \\ of 89.00 is less than average detracted from this prediction.\end{tabular} \\ \hline
\multicolumn{2}{c}{Random Forest} \\ \hline
 & Co-Optimised \\ \hline
Dataset 1 & \begin{tabular}[c]{@{}l@{}}The customer is predicted to stay with the company because their monthly charges of 165.83 are less than average. \\ The fact that their contract category is not Month-to-month, their contract category is not One year, their payment \\ method category is not Electronic Check, their Total Charges of 370.50 are less than average detracted from this \\ prediction.\end{tabular} \\ \hline
Dataset 2 & \begin{tabular}[c]{@{}l@{}}The individual is predicted to be a smoker because their gender is M, their waist of 98.00 is more than average,\\  their age of 40.00 is less than average, and their fasting blood sugar of 106.00 is more than average. The fact that \\ their ldl of 106.00 is less than average detracted from this prediction.\end{tabular} \\ \hline
 & Traditional \\ \hline
Dataset 1 & \begin{tabular}[c]{@{}l@{}}The customer is predicted to stay with the company because their tenure of 2093.76 is less than average and their \\ monthly charges of 165.83 are less than average, and their total charges of 370.50 are less than average. The fact that \\ their internet service category is not Fiber optic, and their contract category is not Month-to-month detracted from \\ this prediction.\end{tabular} \\ \hline
Dataset 2 & \begin{tabular}[c]{@{}l@{}}The individual is predicted to be a smoker because their gender is M, their gtp of 127.00 is more than average, \\ their age of 40.00 is less than average, and their alt of 24.00 is less than average. The fact that their triglyceride of \\ 89.00 is less than average detracted from this prediction.\end{tabular} \\ \hline
\end{tabular}%
}
\end{table}

\section{Limitations and Future Work}
Despite the best efforts of the authors, several limitations of this study can be conceived, and possible avenues for future work are proposed on this basis. The categories of limitations can be divided into data collection, co-optimisation, explanation evaluation, and possible improvements and extensions for each venue, which are detailed below. 

\subsection{Data Collection}
This was the first study to implement the FUS to collect understandability scores and analyse their attributes. Based on the distribution and variation of the scores, several conclusions can be drawn. First, the limitations of the small dataset in a proof-of-concept study need to be acknowledged. As a result, the variance across feature ratings is relatively large, and the representativeness of the scores is uncertain. As such, future research should implement the methodology established in this study on larger FUS datasets. As this was the first instance of FUS usage, several recommendations for future understandability score collection can be made based on observed response behaviour. The first encountered problem was fence-sitting, where users overwhelmingly selected neutral scores, resulting in relatively narrow ranges in understandability. To combat this, future research can include clarifying information, attention checks, and financial incentives to improve the quality of collected responses. Additionally, some features that appeared simple at face value proved relatively difficult for users, likely due to confusion about their measurement scales. As such, future research may wish to provide clarifying information about measurements when scales are deemed ambiguous. Finally, data collection through snowball sampling may introduce sampling biases depending on the domain and network of the researcher, and future research should aim to use random sampling practices.

The current research was limited to two datasets in relatively accessible domains. The collected explainability scores were distributed within a small range centred in the middle of the scale, with no features being considered excessively easy or difficult. While other domains (e.g. medicine or finance) may introduce less understandable features, the ratings are collected in the context of the given domain. As such, users will complete ratings relative to other items in the given dataset, and ratings are expected to remain somewhat centred within a given domain. 

\subsection{Co-Optimisation}
The current study took a filter approach to the co-optimisation of understandability and accuracy. While this permitted the maintenance of high accuracy levels while promoting understandable features at low computational cost, future research may wish to ascertain whether wrapper or embedded methods can do so more effectively. In addition, the current study treats understandability as a spectrum. However, it may be of interest to discretise the understandability values into 'levels' to ascertain how this affects explanation quality. An example is that the co-optimised explanations for Dataset 2 still contain two specialised medical terms, and it may be of interest to assess whether the presence or quantity of difficult features is more impactful for explanation understanding. 

In addition to implementing other cost-sensitive feature selection methods, future research may wish to consider the implementation of more complex machine learning models. This study tested Random Forests, Decision Trees and SVMs for classification, but it may be of interest to assess whether the accuracy-understandability trade-off can be minimised with more complex models such as Gradient Boosting or small neural networks designed for tabular data, such as TabNet \cite{arik2021tabnet}. While these networks carry some implicit challenges, such as decreased transparency, the integration of understandability and usage of ad-hoc explainability methods should combat these problems. 

Additionally, this research assessed the feature selection and classification performance of all models using only a stratified train-test split. Future research may wish to investigate the stability of selected features across random seats and to implement cross-validation or bootstrapping to assess their effects on both the selected features and changes in understandability scores. This would also permit the computation of the statistical significance of the changes.

Another possible extension of the current research is the introduction of a balancing coefficient adjusting the prioritisation of understandability and accuracy. While the proposed filter-based method does not allow for this, an embedded co-optimisation may enable it by introducing an understandability coefficient into the loss function, which adjusts the impact of understandability scores during model optimisation. This would allow prioritising understandability and accuracy scores by domain and target audience, thereby enabling a more flexible integration of understandability into the machine learning training process. 

\subsection{Explanation Evaluation}
Based on the generated explanations, it is notable that features that require one-hot encoding may be more difficult to understand, depending on the category labels and the user's understanding of the encoding logic. As such, it may be of interest to weigh measurement-scale and category-understanding questions more heavily in the computation of the understandability score for features that require one-hot encoding. This would ensure that categories included in the final explanation do not decrease understandability. Alternatively, it may be worthwhile to explore the concatenation of one-hot encoded variables into the parent feature before the creation of explanations. However, this would decrease the explanation's faithfulness, which is a trade-off that may not be desirable depending on the work's context and domain. However, the most important next step is to collect user feedback to evaluate the change in explanation quality resulting from the co-optimisation procedure. While explanations were assessed through changes in understandability cost and at face value in the current study, user feedback is necessary to ensure that the quality improvements are both present and effective in the end-user base. Ideally, explanations should be evaluated using an assessment scale and compared with traditional explanations to determine which users prefer. The authors are currently implementing a validation study to evaluate changes in quality. To this end, future studies would also be of interest to develop and validate a scale to measure the understandability and quality of textual explanations. While similar scales assessing various dimensions of explanation quality exist, none have been found to specifically evaluate the quality of textual explanations. 

\section{Conclusion}
This proof-of-concept study aimed to implement the FUS on two datasets, analyse the score distribution, propose a co-optimisation technique and evaluate the resulting explanations. The FUS scores showed a limited range of feature understandability across both datasets, and reasons as well as possible solutions for this were proposed. The accuracy in co-optimised models decreased somewhat compared to the traditional setting. However, the co-optimised models also successfully promoted the use of understandable features, decreasing the mean and total understandability cost and generating more understandable explanations at face value. These findings highlight that improved explanations can be generated at low accuracy cost through the promotion of understandable features. However, further research collecting user feedback on the explanations' quality is necessary to assess whether there is a measurable change in quality.

\begin{credits}
\subsubsection{\ackname} This work was conducted with the financial support of Research Ireland - Taighde Éireann, under Grant Nos. 18/CRT/6223 and 12/RC/2289-P2, the latter being co-funded under the European Regional Development Fund. For the purpose of Open Access, the author has applied a CC BY public copyright licence to any Author Accepted Manuscript version arising from this submission.

\subsubsection{\discintname}
The authors have no competing interests to declare that are
relevant to the content of this article.
\end{credits}

\bibliography{references}




% LNCS Homepage, \url{http://www.springer.com/lncs}, last accessed 2023/10/25
% \end{thebibliography}
\end{document}